\documentclass{iopjournal}

\usepackage{graphicx}
\usepackage{amsmath}
\usepackage{amssymb}
\usepackage{dcolumn}
\usepackage{bm}
\usepackage[mathlines]{lineno}
\usepackage[normalem]{ulem}
\usepackage[abs]{overpic}
\usepackage{color}
\usepackage{rotating}
\usepackage{enumerate}
\usepackage{siunitx}
\usepackage[section]{placeins}
\usepackage{physics}
\usepackage{comment}
\usepackage{tikz, color}
\usepackage{hyperref}
\usepackage{cite}
\begin{document}

\articletype{Article type} 

\title{A mathematical justification to apply the secular approximation to the Redfield equation}

\author{Niklas J. Jung$^{1,2,*}$\orcid{0009-0003-3417-2281}, Francesco Rosati$^{1,2}$\orcid{0009-0002-3667-4202}, Gabriel L. Rath$^{1,2}$\orcid{0009-0008-2068-2789},
Frank K. Wilhelm$^{1,2}$\orcid{0000-0003-1034-8476},
Peter K. Schuhmacher$^{3}$\orcid{0000-0003-1232-4363}}

\affil{$^1$Theoretical Physics, Saarland University, Campus, 66123 Saarbr\"{u}cken, Germany}

\affil{$^2$Institute for Quantum Computing Analytics (PGI-12), Forschungszentrum J\"ulich, 52425 J\"ulich, Germany}

\affil{$^3$Department of High Performance Computing, Institute of Software Technology, German Aerospace Center (DLR),  Rathausallee 12, 53757 Sankt Augustin, Germany}

\affil{$^*$Author to whom any correspondence should be addressed.}

\email{n.jung@fz-juelich.de}

\begin{abstract}

Quantum master equations are widely used to describe the dynamics of open quantum systems. All these different master equations rely on specific approximations that may or may not be justified. Starting from a microscopic model, applying the standard Born and Markov approximations results in the Redfield equation that does not guarantee to preserve positivity. The latter is typically achieved by additionally applying the secular approximation resulting in a quantum master equation in Lindblad form.
There are other ways to obtain an equation in  Lindblad form, one of which is the
recently proposed Universal Lindblad Equation.
It has been shown that it is in the same equivalence class of approximations as the Redfield master equation although avoiding the heuristic secular approximation \cite{RudnerNathan}. In this work, we prove that the solutions of the master equation obtained by applying the secular approximation are also obtained by an approximation of the same order as the one performed to obtain the Redfield equation. We hereby provide a mathematical justification for the secular approximation.
We show that the result of applying the secular approximation is obtained naturally by applying a self-consistency argument. This shows that the resulting master equation is also in the same equivalence class of approximations as the Redfield master equation and the Universal Lindblad Equation. We furthermore give numerical considerations about the accuracy of the secular master equation compared to the Universal Lindblad equation.
\end{abstract}

\newpage

\section{Introduction}

The characterization and description of open quantum systems is an important problem in physics that has been studied for a long time \cite{Breuer2006}. It is relevant for many fields, including atomic, molecular, and optical physics. Its significance extends especially to the field of quantum information processing \cite{Nielsen_Chuang}, where describing open quantum systems helps to understand and develop control techniques \cite{Koch_2022}. Modern quantum technologies thus require a precise description of open quantum systems.

A powerful tool in the study of these open quantum systems are quantum master equations. Their validity, however, is often unclear, as it typically relies on uncontrolled approximations \cite{Breuer2006,soret2025}: starting from the exact Liouville-Von Neumann equation, one performs the standard Born (assuming weak coupling between system of interest and its environment) and Markov approximations (assuming short bath memory) to derive the Redfield equation \cite{Redfield1957}.

A more general approach to derive a master equation to describe an open quantum system is the Nakajima-Zwanzig formalism \cite{Zwanzig,Nakajima}. In its most general form, it leads to an exact but hard (or even impossible) to solve master equation. This equation is usually truncated to get an approximate but more easily solvable master equation, although the validity of these approximations is not always clear \cite{Stockburger}. The aforementioned Redfield equation emerges as the lowest order approximation of the Nakajima-Zwanzig equation
\cite{Breuer2006}. Naturally, considering only the lowest order will lead to errors. A discussion of its limitations as well as a treatment of higher orders can be found in \cite{Crowder_2024}.

However, the Redfield equation is not a Lindblad master equation, which in particular means that it does not guarantee to preserve positivity \cite{Lindblad1976,Gorini1976,Trushechkin,Lindblad:1975ef}. The traditional way to achieve Lindblad form is through another approximation called the secular approximation. This approximation is usually performed heuristically by neglecting "fast oscillating terms", which means neglecting oscillating terms containing a frequency difference $\omega - \omega'$ if $\omega \neq \omega'$. The resulting equation is sometimes called the Quantum Optical Master Equation (QOME), which we will call it in this work. It offers a range of advantages compared to the Redfield Equation \cite{Breuer2006,SoretSecular}. First, being in Lindblad form, it generates a completely positive quantum-dynamical map. Furthermore, a master equation in Lindblad form can alternatively be solved numerically by stochastic evolution of pure states \cite{Breuer2006, PhysRevLett1}.\\
An alternative and novel way to achieve a quantum master equation in Lindblad form, which is based on the Redfield equation too, is called the Universal Lindblad Equation (ULE)  \cite{RudnerNathan}. It does not rely on the heuristic secular approximation. Additionally, it does not require knowledge of the system eigenstates to obtain the Lindblad operators, as opposed to the QOME. This makes it more generally applicable. More phenomenological approaches to achieving a beyond-secular, Lindblad type master equation have been pursued as well \cite{Kir_anskas_2018}. However, while these approaches, in many cases, lead to accurate solutions, they lack rigorous mathematical foundation.

In this work we show that the solutions of the QOME are valid up to the same order of approximation as the corresponding solutions of the Redfield equation it has been derived from. For this purpose we present a novel, systematic derivation of the QOME, resulting in the exact same result as applying the secular approximation. This shows that the Quantum Optical Master Equation belongs to the same equivalence class of Markov approximations as the Redfield equation and the Universal Lindblad Equation. In the process, we also provide a framework for application of the secular approximation, which is usable even in higher orders of the Nakajima-Zwanzig equation.
We furthermore give analytical as well as numerical considerations as to under which conditions the QOME outperforms the ULE with respect to accuracy.

We present the rigorous derivation of the QOME in section \ref{sec:LME}. We start by giving a brief overview of the Nakajima-Zwanzig approach to derive the Nakajima-Zwanzig Master Equation in section \ref{sec:RME}. We show how this equation is approximated to the Redfield equation and give an explicit form for the Redfield tensor. We write down the formal solution of the Redfield equation. This enables us to derive the QOME from the Redfield equation in section \ref{sec:secular} while showing that none of the further neglected terms are of higher order than those already neglected to get to the Redfield equation. This proves that the solution of the thus obtained QOME is valid up to the same order of approximation as the solution of the Redfield Equation. We furthermore provide an analytical consideration in section \ref{sec:comp} which indicates that the accuracy of the QOME compared to the ULE is dependent on system parameters. In particular we show strong indications that the temperature of the bath determines the accuracy of the ULE compared to the QOME, with higher temperatures favoring the ULE.

To verify this claim, in section \ref{sec:numerics}, we compare the accuracy of the solutions of the QOME and the ULE numerically in certain cases where the Redfield equation retains valid results -- and it can henceforth serve as a benchmark. We observe that, for low temperatures, the numerical solution of the QOME shows a higher similarity to that of the Redfield equation than that of the ULE. However, for higher tempreatures the ordering reverses, and the solutions of the ULE become more accurate.  
We also know from previous results \cite{Davidovic2020completelypositive} that there are other regimes where the QOME in fact performs worse than ULE-type equations. This discrepancy is discussed briefly. However, the general conditions under which the QOME outperforms ULE-type equations are still an open question.

\section{The Lindblad Master Equation} \label{sec:LME}

\label{sec:Lindblad}
{In the theory of open quantum systems, the state of a system at time $t$ is described by a Hermitian, positive semi-definite matrix $\hat{\rho}(t)$ with ${\rm
Tr}[\hat{\rho}(t)]=1$ called the density matrix \cite{Breuer2006}. To describe its dynamics, it is therefore desirable to use a quantum master equation that preserves these properties for any $t$, that is, it has to be completely positive and trace-preserving. As proven by Lindblad in 1976 \cite{Lindblad1976}, such a master equation can always be written in the form} 

\begin{eqnarray}
\label{LindbladEquation}
\frac{{\rm d}}{{\rm
d}t}\hat{\rho}(t)= -i\left[\hat{H},\hat{\rho}(t)\right]
+\sum_{k}\gamma_k\left(\hat{L}_k\hat{\rho}(t)\hat{L}_k^{\dagger}-\frac{1}{2}\left\{\hat{L}_k^{\dagger}\hat{L}_k,\hat{\rho}(t)\right\}\right).
\end{eqnarray}

{Here,} the time evolution of the density matrix $\hat{\rho}(t)$ is fully determined by a unitary part, generated by the Hamiltonian $\hat{H}$, and a dissipative part given by the Lindblad operators $\hat{L}_k$. Eq. \eqref{LindbladEquation} has been applied to many tasks within a wide range of research fields, from quantum optics~\cite{Gardiner2004,Buchheit2016} and atomic physics~\cite{Metz2006,Jones2018,Cohen1992} to quantum information~\cite{Whitfield2010,Cuevas2012,Briegel2014,Govia2017,Schuhmacher2021} and many more.\\
{Although the form of the desired master equation is known, it remains an open question how to derive such a master equation accurately based on a microscopic model in general. For example, Ref.~\cite{Buchheit2016} illustrates well how the use of an inaccurate Lindblad equation can lead to discrepancies between experiments and the theory describing them. Hence, given a quantum system of interest described by the Hamiltonian $\hat{H}_{\rm S}$ that couples to an environment described by the Hamiltonian $\hat{H}_{\rm B}$, how do the correct Lindblad operators  $\hat{L}_k$ in Eq.~\eqref{LindbladEquation} look like to accurately capture the system dynamics?}

\subsection{Preliminaries: Derivation of the Redfield Equation} \label{sec:RME}

We consider a total quantum system described by the Hamiltonian
\begin{eqnarray}
\label{GeneralHamiltonian}
\hat{H}=\hat{H}_{\rm S}+\hat{H}_{\rm B}+g\hat{H}_{\rm int}.
\end{eqnarray}
Here, $\hat{H}_S$ and $\hat{H}_B$ denote the Hamiltonians corresponding to the quantum system of interest S and the bath B respectively, while the interaction between system and bath with dimensionless coupling strength $g \ll 1$ is described by the Hamiltonian
\begin{eqnarray}
\label{DecompositionInteractionHamiltonian}
\hat{H}_{\rm int}=\sum_\alpha \hat{S}_\alpha \otimes\hat{B}_\alpha
\end{eqnarray}
that can be decomposed in operators $\hat{S}_{\alpha}$ and $\hat{B}_{\alpha}$ acting respectively on the system and on the bath only. We consider $\dim(\hat{H}_S)=z\in\mathbb{N}$ in the following.

The standard way to derive a quantum master equation in Lindblad form is to derive the Redfield master equation first, and to apply the secular approximation afterwards \cite{Redfield1957,Breuer2006}. In the following, we show that we can achieve the same result by only dropping terms which are of same order than those we already dropped to derive the Redfield equation.

There are plenty of ways to derive the Redfield master equation, like the phenomenological approach by Redfield himself by applying the standard Born and Markov approximations \cite{Redfield1957}, or the Bogoliubov method \cite{Trushechkin,Bogoliubov1946}. Here, we use the systematic derivation of the Redfield master equation based on projection operator techniques as presented in Ref.~\cite{Breuer2006}, chapter 9. 

We arrive at a master equation in the interaction picture which reads

\begin{eqnarray} 
\label{MarkovianBlochRedfieldEquation} 
\frac{{\rm d}}{{\rm d}t}\hat{\tilde{\rho}}_{\rm S}(t) \,=
-g^2 \int_0^{t} {\rm d} s
\left({\rm Tr}_{\rm B}\left [ \hat{H}_{\rm int}(t) \right.,\left.\left[\hat{H}_{\rm int}(t-s),\hat{\tilde{\rho}}_{\rm S}(t)\otimes\hat{\tilde{\rho}}_{\rm B}\right ]\right ]\right)
+\, \mathcal{O}\left(g^4\right),
\label{eq:NakajimaZwanzig}
\end{eqnarray}
where we have denoted the reduced density operator of the system and bath by $\hat{\rho}_S$ and $\hat{\rho}_B$ respectively, and their interaction picture counterparts as $\hat{\tilde{\rho}}_S$ and $\hat{\tilde{\rho}}_B$.

By dropping the terms of $\mathcal{O}(g^4)$, Eq.~\eqref{MarkovianBlochRedfieldEquation} becomes the well-known Redfield equation. A treatment of higher orders of this expansion can be found in \cite{Crowder_2024}.

The Redfield Equation generally can not be written in Lindblad form in a way that all $\gamma_k$ are positive. In particular, Eq.~\eqref{MarkovianBlochRedfieldEquation} does not guarantee to preserve positivity \cite{ZollerGardiner}. 

Let now $\varepsilon_k$ be the eigenfrequencies of $\hat{H}_{\rm S}$ such that $\hat{\Pi}(k)$ projects onto the subspace indexed by $k$. We can write Eq.~\eqref{MarkovianBlochRedfieldEquation} in the Schr\"odinger picture using the definitions
\begin{equation}
    \hat{S}_{\alpha}(\omega)=\sum_{\varepsilon_k-\varepsilon_{k'}=\omega}\hat{\Pi}(k)\hat{S}_{\alpha} \hat{\Pi}(k')
\end{equation}
and
\begin{equation}
    \Gamma_{\alpha\beta}(\omega)=g^2 \int_{0}^{t}e^{i\omega s}\mathcal{B}_{\alpha\beta}(s){\rm d}s
\end{equation}
with the bath correlation functions
\begin{eqnarray}
\mathcal{B}_{\alpha\beta}(s)={\rm Tr}_{\rm B}\left[\hat{B}_\alpha(t)\hat{B}_\beta(t-s)\hat{\rho}_{\rm B}\right]
=\left\langle\hat{B}_\alpha (s)\hat{B}_\beta (0)\right\rangle,     
\end{eqnarray}
resulting in
\begin{equation}
\label{RedfieldEquationTensor}
\frac{{\rm d}}{{\rm d}t}\hat{\rho}_{\rm S}(t)=-i\left[\hat{H}_{\rm S},\hat{\rho}_{\rm S}(t)\right]+\hat{R}\hat{\rho}_{\rm S}(t),
\end{equation}
with the Redfield tensor

\begin{eqnarray}
\label{Rrho}
  \nonumber\hat{R}\hat{\rho}(t)&=\displaystyle{\sum_{\omega,\nu}\sum_{\alpha,\beta}}\Gamma_{\alpha\beta}(\omega)\left(\hat{S}_{\beta}(\omega)\hat{\rho}(t)\hat{S}^{\dagger}_{\alpha}(\nu)-\hat{S}^{\dagger}_{\alpha}(\nu)\hat{S}_{\beta}(\omega)\hat{\rho}(t) \right) + \text{H.c.},
\end{eqnarray}

where we dropped the subscript S to shorten the notation, since from now on the considered density matrix is always the one of that system of interest S.
In an eigenbasis of the system Hamiltonian, Eq.~\eqref{RedfieldEquationTensor} written in component form becomes
\begin{eqnarray}
\label{RedfieldEquationComponents}
	\frac{{\rm d}}{{\rm d}t} \rho_{ab}(t) = 
	-i \omega_{ab} \rho_{ab}(t) +
	\sum_{j,k} R_{abjk} \rho_{jk}(t),
\end{eqnarray} 

where $\omega_{jk} = \varepsilon_j - \varepsilon_k$ denote the Bohr frequencies. By definion the matrix elements of $\hat{R}$ satisfy $R_{abjk}=\mathcal{O}\left(g^2\right)$ and
\begin{equation}
    \left( \hat{R} \rho(t) \right)_{ab} = 
    \bra{a} \hat{R} \rho(t) \ket{b} =
    \sum_{j,k} R_{abjk} \rho_{jk}.
    \label{eq:coefficientcomparisonsum}
\end{equation}

We first explicitly calculate the term
\begin{eqnarray}
    \bra{a} \hat{S}_\beta(\omega) \hat{\rho}(t) \hat{S}_\alpha^\dagger(\nu) \ket{b}
    &&= 
	\sum_{\stackrel{n,m,}{\omega_{nm} = \omega}} 
	\sum_{\stackrel{p,l,}{\omega_{pl} = \nu}} 
    \bra{a}\ket{n} \bra{n} \hat{S}_\beta \ket{m} \bra{m} \rho(t) 
    \ket{l} \bra{l} \hat{S}_\alpha \ket{p} \bra{p} \ket{b} \nonumber \\
    &&= 
    \sum_{\stackrel{m,}{\omega_{am} = \omega}} 
	\sum_{\stackrel{l,}{\omega_{bl} = \nu}} 
    \bra{a} \hat{S}_\beta \ket{m} \bra{l} \hat{S}_\alpha \ket{b} \rho_{ml}(t).
\end{eqnarray}
We can do the same for the other terms in Eq.~\eqref{Rrho} and then compare the coefficients in the sum in Eq.~\eqref{eq:coefficientcomparisonsum}. 
We observe that the elements of the Redfield tensor can be rewritten as
\begin{eqnarray}
\label{RedfieldtensorKompnenten}
    R_{abjk} =
	   \sum_{\omega,\nu} \left( 
			\mathcal{S}^{\left({1}\right)}_{ab,\omega\nu,jk} - 
			\mathcal{S}^{\left({2}\right)}_{ab,\omega\nu,jk}
            + 
			\mathcal{S}^{\left({3}\right)}_{ab,\omega\nu,jk} -
			\mathcal{S}^{\left({4}\right)}_{ab,\omega\nu,jk} \right),
\end{eqnarray}
using the definitions
\begin{align}
\label{definitionS}
    \mathcal{S}^{\left({1}\right)}_{ab,\omega\nu,jk} &= 
        \sum_{\alpha,\beta} 
        \Gamma_{\alpha\beta}(\omega)
		\sum_{\stackrel{m}{\omega_{am} = \omega}} 
		\sum_{\stackrel{l}{\omega_{bl} = \nu}} 
		\langle a| \hat{S}_\beta | m\rangle
		\langle l| \hat{S}_\alpha | b\rangle 
		\delta_{mj} \delta_{lk} \notag \\ 
    \mathcal{S}^{\left({2}\right)}_{ab,\omega\nu,jk} &= 
        \sum_{\alpha,\beta} 
        \Gamma_{\alpha\beta}(\omega)
		\sum_{\stackrel{m}{\omega_{ma} = \nu}} 
		\sum_{\stackrel{l}{\omega_{ml} = \omega}} 
		\langle a| \hat{S}_\alpha | m\rangle
		\langle m| \hat{S}_\beta | l\rangle 
		\delta_{lj} \delta_{bk} 
\end{align}
and similar definitions for $\mathcal{S}^{\left({p}\right)}_{ab,\omega\nu,jk}, p \in \{3,4\}$.

In the next section, we show that there is a systematic method to perform the secular approximation. We will use this method to simplify Eq.~\eqref{RedfieldtensorKompnenten}. This will yield an equation that is equivalent to the Redfield equation in approximation order, but is a Lindblad master equation. 

\subsection{Self-Consistent Analysis of the Solutions to the Redfield Equation} \label{sec:secular}

Note that to get to the Redfield equation, we neglect all terms of order $\mathcal{O}\left(g^4\right) = \mathcal{O}\left(R^2\right)$ in Eq.~\eqref{MarkovianBlochRedfieldEquation}. Any further approximation which only discards terms of this order will therefore be self-consistent and therefore equivalent in approximation order. As we show in the following, the secular approximation emerges as a direct result of this self-consistency argument.

We start by vectorizing Eq.~\eqref{RedfieldEquationComponents}, which reduces the problem of solving the differential equation to solving the characteristic equation of a matrix. We show that we can neglect certain terms in this characteristic equation. This leads to the double sums in Eq.~\eqref{Rrho} reducing to a single sum over one parameter, which corresponds to the usual way of applying the secular approximation \cite{Breuer2006}.

First, we vectorize Eq.~\eqref{RedfieldEquationComponents}. For the indices $i, j,$ $k, l$ $\in 
\{n \in \mathbb{N}, n < z \}$ in matrix form we define
\begin{eqnarray}
	c = iz + j, d = kz + l \in \{n \in \mathbb{N}, n < z^2 - 1 \},
\end{eqnarray}
so for example in the case of a two-level system the vectorized density matrix can be written as
\begin{equation}
    \hat{\rho}(t)=\left(\begin{array}{rr} \rho_{00}(t)     & \rho_{01}(t) \\
    \rho_{10}(t)     & \rho_{11}(t)
    \end{array}\right)\longrightarrow \underline{\rho}(t)=\left(\begin{array}{c}
       \rho_{00}(t)  \\
       \rho_{01}(t)  \\
       \rho_{10}(t)  \\
       \rho_{11}(t)  
    \end{array}\right).
\end{equation}
For the density matrix and the frequencies in the Redfield Eq.~\eqref{RedfieldEquationComponents}) we get
\begin{eqnarray}
\label{Notation}
	\rho_c = \rho_{ij} \textnormal{ and } \omega_c = \omega_{ij}.
\end{eqnarray}
We then define the matrices $\mathcal{H}$ with
\begin{eqnarray}
	 \mathcal{H} = 
	-i\cdot\textnormal{diag}\left( \omega_0, \omega_1, \dots, \omega_{z^2 - 1} \right)
\end{eqnarray}
and $\mathcal{R}$ with
\begin{eqnarray}
	\mathcal{R}_{cd} = R_{ijkl}.
\end{eqnarray}
The Redfield Eq.~\eqref{RedfieldEquationComponents} can be written as
\begin{eqnarray}
	\frac{{\rm d}}{{\rm d}t}\underline{\rho}(t) = \left(\mathcal{H} + \mathcal{R}\right)\underline{\rho}(t).
\end{eqnarray}
This system of ordinary differential equations can be solved by the eigenanalysis method. The general solution is given by \cite{bookODEs}
\begin{align}
    \underline{\rho}(t) = \sum_j c_j e^{s_j t} \underline{v}_j,
\end{align}
where $s_j$ and $\underline{v}_j$ are the eigenvalues and eigenvectors of $\mathcal{H} + \mathcal{R} =: \mathcal{M}$. Plugging in $\underline{\rho}(0) = \underline{\rho}_0$ gives the unique solution to the initial value problem.

The characteristic equation is given by $\det\left(\mathcal{A}(s)\right)=0$ of $\mathcal{M}$, where $\mathcal{A}(s)=-s \mathbb{I} + \mathcal{H} + \mathcal{R}$. 
Note again that in the derivation of Eq.~\eqref{RedfieldEquationComponents}, we neglected terms of order $\mathcal{O}(g^4)$. Therefore, if we drop these in the characteristic equation, the solution will be of the same order of approximation. In the following, our goal is to show that if we do that, we in fact get back the QOME.

For the following considerations, we distinguish between two types of eigenvalues. Since $\mathcal{H}$ is diagonal, the eigenvalues $s_j$ of $\mathcal{M}$ have to be of the form $s_j=i\omega_j+\delta s_j $, with $\delta s_j = \mathcal{O}(\mathcal{R})$. Note that $\mathcal{O}(\mathcal{R}) = \mathcal{O}(g^2)$ as the Redfield tensor is of order $\mathcal{O}(g^2)$. If $\omega_j=0$, then we call $s_j$ an incoherent pole. If $\omega_j\neq0$, then we call $s_j$ a coherent pole.

We first consider all the diagonal elements of $\mathcal{A}$ that are of order $\mathcal{O}(g^2)$. We define the set
\begin{equation}
    D=\left\{\left.d \in \left\{0,\dots,z^2-1\right\} 
		\right| \mathcal{A}_{dd} \in \mathcal{O}(g^2)\right\}.
\end{equation}
This is a finite set. We call its number of elements $n_D : =\# D$ and index its elements by $d_0,\ldots d_{n_D-1}$. Note that if $c\notin D$, then $\mathcal{A}_{cc}=\mathcal{O}(g^0)$. Furthermore, we have $\mathcal{A}_{cd}\in\mathcal{O}(g^2)$ for $c \neq d$, since $\mathcal{H}$ is diagonal. Now, we compute the determinant using the Leibniz formula. It follows 
\begin{align}
		\nonumber\det\left({\mathcal{A}\left({s}\right)}\right) &= 
        \displaystyle{\sum_{\tau \in S_{z^2}}
		{\rm sgn}\left({\tau}\right)\prod_{i=0}^{z^2-1}
		\mathcal{A}_{i,\tau\left({i}\right)}\left({s}\right)}  \\
		&= \displaystyle{\underbrace{\left(
		\prod_{\stackrel{i=0}{i \notin D}}^{z^2-1}
		\mathcal{A}_{ii}\left({s}\right) \right)}_{\mathcal{O}(g^0)}
        \underbrace{
		\sum_{\tau \in U_{D}} {\rm sgn}\left({\tau}\right)
		\prod_{k \in D} \mathcal{A}_{k,\tau\left({k}\right)}\left({s}\right)}_
        {{\mathcal{O}\left({g^{2n_D}}\right)}}} + \mathcal{O}\left({g^{2(n_D+1)}}\right),
		\label{detUz2}
	\end{align}
where $U_{D}= \left\{{\tau \in S_{z^2}, \tau\left({i}\right) = i \; 
		\forall \; i \notin D}\right\}$. 
        By definition, $\mathcal{A}_{ii}$ is of order $\mathcal{O}(1)$ for $i \notin D$. For all other $c,d$ we have by definition $\mathcal{A}_{cd} \in \mathcal{O}\left({g^2}\right)$. The first term on the right hand side of Eq.~\eqref{detUz2} is therefore of order $\mathcal{O}\left({g^{2n_D}}\right)$. 
        Furthermore, the $A_{ii}$ for $i \notin D$ are the only elements of $\mathcal{A}$ which are not of order $\mathcal{O}\left({g^2}\right)$. This means that, except for the aforementioned first term, all other terms contained in the determinant are at least of order $\mathcal{O}\left({g^{2(n_D+1)}}\right)$.
        
        Let us consider the equation 
        \begin{align}
            \det\left(\mathcal{A}(s)\right)=0.
        \end{align}
        We now apply the self-consistency argument. To derive the Redfield equation from the Nakajima-Zwanzig equation (Eq.~\eqref{eq:NakajimaZwanzig}), we only kept the terms in the equation that are of lowest order in $g$, so terms of order $\mathcal{O}(g^2)$. All terms of higher order, namely those of order $\mathcal{O}(g^4)$ were dropped. For the determinant, as illustrated above, the terms of lowest order in $g$ are of order $\mathcal{O}(g^{2n_D})$ Therefore, to apply this approximation consistently here, we drop all terms that are of order $\mathcal{O}\left({g^{2(n_D+1)}}\right)$. This leaves us with the equation
        \begin{equation}
        \left(
		\prod_{\stackrel{i=0}{i \notin D}}^{z^2-1}
		\mathcal{A}_{ii}\left({s}\right) \right)
        \sum_{\tau \in U_{D}} {\rm sgn}\left({\tau}\right)
		\prod_{k \in D} \mathcal{A}_{k,\tau\left({k}\right)}\left({s}\right) = 0, 
        \end{equation}
        where we can factor out $\left(
		\prod_{\stackrel{i=0}{i \notin D}}^{z^2-1}\mathcal{A}_{ii}\left({s}\right) \right) \neq 0$ to obtain 

\begin{equation}
   \sum_{\tau \in U_{D}} {\rm sgn}\left({\tau}\right)
		\prod_{k \in D} \mathcal{A}_{k,\tau\left({k}\right)}\left({s}\right) =:
		\det\left({\tilde{\mathcal{A}}\left({s}\right)}\right) = 0.
\end{equation}
Here, we defined the $n_D \times n_D$-matrix $\tilde{\mathcal{A}}$ as
\begin{equation}
    \tilde{\mathcal{A}} = \left({\tilde{\mathcal{A}}_{ij}\left({s}\right)}\right)_{i,j \in
			\left\{{0,\dots,n_D-1}\right\}}
		\textnormal{ with } \tilde{\mathcal{A}}_{ij}\left({s}\right) =
		\mathcal{A}_{d_id_j}\left({s}\right).
		\label{eq18}
\end{equation}
We can visualize this process by the matrix represented in fig. \ref{fig:matrixdemo}. All the nondiagonal terms of this matrix are of order $\mathcal{O}(g^2)$. Only on the diagonal, we can have terms of order $\mathcal{O}(g^0)$. This means that the term of lowest order in $g$ has to be the product of all diagonal elements. Furthermore, any permutation that leads to a term of the same order has to only swap out diagonal elements of order $\mathcal{O}(g^2)$. All of these permutations are contained in the set $U_D$. 

Hence, we can restrict ourselves to only calculating the determinant of the reduced matrix $\tilde{\mathcal{A}}$. This is equivalent to the fact that, instead of $\mathcal{R}$, we only consider the reduced matrix 
\begin{align}
    \tilde{\mathcal{R}} = \left({\tilde{\mathcal{R}}_{ij}\left({s}\right)}\right)_{i,j \in
			\left\{{0,\dots,n_D-1}\right\}}
		\textnormal{ with } \tilde{\mathcal{R}}_{ij}\left({s}\right) =
		\mathcal{R}_{d_id_j}\left({s}\right).
\end{align}

The elements contained in the reduced matrix $\tilde{R}$ depend on the set $D$, which differs for coherent and incoherent poles. For coherent poles, we have
\begin{equation}
 \mathcal{A}_{dd} = -i\omega_c + \delta_s + i\omega_d + R_{dd} \in
	\mathcal{O}(g^2)\Leftrightarrow\omega_d=\omega_c,   
\end{equation}
or equivalently
\begin{equation}
    D = \left\{{d \in \left\{{0,\dots,z^2-1}\right\}; \; \omega_d \neq 0; \; 
			\omega_d = \omega_c}\right\}.
\end{equation}
For incoherent poles, we have $s = \delta_s \in
	\mathcal{O}(g^2)$, such that
\begin{equation}
\mathcal{A}_{dd} = \delta_s + i\omega_d + R_{dd} \in \mathcal{O}(g^2)\Leftrightarrow\omega_d = 0,
\end{equation}
hence
\begin{equation}
   D = \left\{{d \in \left\{{0,\dots,z^2-1}\right\}; \; \omega_d = 0}\right\}. 
\end{equation}

\begin{figure}
\centering
\begin{equation}
    \begin{pmatrix}
        \mathcal{O}(g^2) & \cdots & \square & \cdots & \square & \cdots &
        \square & \cdots \\
        \vdots & \ddots & \vdots &  & \vdots & & \vdots \\
        \square & \cdots & \mathcal{O}(g^2) & \cdots & \square & \cdots &
        \square & \cdots \\
        \vdots &   & \vdots & \ddots & \vdots & & \vdots \\
        \square & \cdots & \square & \cdots & \mathcal{O}(g^2) & \cdots &
        \square & \cdots \\
        \vdots &   & \vdots &   & \vdots & \ddots & \vdots \\
        \square & \cdots & \square & \cdots & \square & \cdots &
        \mathcal{O}(g^2) & \cdots \\
        \vdots &   & \vdots &   & \vdots & & \vdots & \ddots 
    \end{pmatrix} \notag
\end{equation}    
\caption{Illustration of the simplification process for the matrix. Only permutations that swap $\mathcal{O}(g^2)$-elements on the diagonal are of lowest order in $g^2$. This means we only have to consider the determinant of the reduced matrix $\tilde{\mathcal{A}}$, which consists of the $\mathcal{O}(g^2)$-elements together with the squares in the figure.}
\label{fig:matrixdemo}
\end{figure}

In both cases, the remaining matrix elements $\tilde{\mathcal{R}}_{ij}$ satisfy $\omega_i=\omega_j$. Going back to the non-vectorised notation $R_{abjk}$, we find
\begin{equation}
\label{SecularResult}
    \omega_{ab} = \varepsilon_a - \varepsilon_b = \omega_c = \varepsilon_j - \varepsilon_k = \omega_{jk}.
\end{equation}
\newpage
We found a condition for $R_{abjk}$ to be negligible, namely that $\omega_{ab} = \omega_{jk}$. This condition we insert in the definition (\ref{definitionS}) of
	$\mathcal{S}^{\left({1}\right)}_{ab,\omega\nu,jk}$. 
We note that by definition for any contribution to $\mathcal{S}^{\left({1}\right)}_{ab,\omega\nu,jk}$, we have
\begin{equation}
    \omega_{aj} = \varepsilon_a - \varepsilon_j = \omega \textnormal{ and }
		\omega_{bk} = \varepsilon_b - \varepsilon_k = \nu,
\end{equation}
which implies using Eq.~\eqref{SecularResult} that
\begin{equation}
    \nu = \varepsilon_b - \varepsilon_k = 
		\left({\varepsilon_a - \omega_c}\right) - \left({\varepsilon_j - \omega_c}\right)
		= \varepsilon_a - \varepsilon_j = \omega.
\end{equation}
We obtain the same result for the other $\mathcal{S}^{\left({p}\right)}_{ab,\omega\nu,jk}$ analogously. This means that the double sum $\sum_{\omega,\nu}$ in Eq. \eqref{RedfieldtensorKompnenten} will simplify to a single sum. 
\subsection{Rediscovering The Secular Approximation}
 To conclude section \ref{sec:secular}, by neglecting terms of higher than the lowest order in $g^2$, we rediscover exactly the standard secular approximation \cite{Breuer2006}. Eq.~\eqref{Rrho} becomes

    \begin{eqnarray}
\label{RrhoSekular}
    \hat{R}\hat{\rho}(t)&=\displaystyle{\sum_{\omega}\sum_{\alpha,\beta}}
    \Gamma_{\alpha\beta}(\omega)\left(\hat{S}_{\beta}(\omega)\hat{\rho}(t)\hat{S}^{\dagger}_{\alpha}(\omega)-\hat{S}^{\dagger}_{\alpha}(\omega)\hat{S}_{\beta}(\omega)\hat{\rho}(t) \right) + \text{H.c.}
\end{eqnarray}

It is known that this equation can be cast into Lindblad form \cite{Breuer2006} by splitting up the coefficients
\begin{equation}
    \Gamma_{\alpha\beta}(\omega) = \frac{1}{2} \gamma_{\alpha\beta}(\omega) + 
    i \Im(\Gamma_{\alpha\beta}(\omega))
\end{equation}
with
\begin{eqnarray}
    \gamma_{\alpha\beta}(\omega)=2\Re\left(\Gamma_{\alpha\beta}(\omega)\right)=g^2\int_{-t}^{t}e^{i\omega s}\mathcal{B}_{\alpha\beta}(s){\rm d}s.
\end{eqnarray}
Diagonalizing the coefficient matrix $\left(\Gamma_{\alpha\beta}(\omega)\right)_{\alpha,\beta}$ yields
\begin{eqnarray}
\label{FastLindblad}
    \hat{R}\hat{\rho}(t)= -i\left[\hat{H}_{\rm LS},\hat{\rho}(t)\right]
    +\sum_{\omega,q}\gamma_{q}(\omega)\left(\hat{S}_q(\omega)\hat{\rho}(t)\hat{S}_q(\omega)^{\dagger}-\frac{1}{2}\left\{\hat{S}_q(\omega)^{\dagger}\hat{S}_q(\omega),\hat{\rho}(t)\right\}\right)\nonumber\\
\end{eqnarray}
with the Lamb shift Hamiltonian
\begin{equation}
   \hat{H}_{\rm LS}=\sum_{\omega,q}\Im\left(\Gamma_q(\omega)\right)\hat{S}_{q}(\omega)^{\dagger}\hat{S}_q({\omega}). 
\end{equation}
We finally arrive at the quantum master equation in Lindblad form
\begin{eqnarray}
&&\frac{{\rm d}}{{\rm d}t}\hat{\rho}(t)= \mathcal{L}_{\text{QOME}}\hat{\rho} = -i\left[\hat{H}_{\rm S} + \hat{H}_{\rm LS},\hat{\rho}(t)\right]\nonumber\\
&&+\sum_{\omega,q}\gamma_{q}(\omega)\left(\hat{S}_q(\omega)\hat{\rho}(t)\hat{S}_q(\omega)^{\dagger}-\frac{1}{2}\left\{\hat{S}_q(\omega)^{\dagger}\hat{S}_q(\omega),\hat{\rho}(t)\right\}\right).\nonumber\\
\label{eq: QOME}
\end{eqnarray}

This is the well-known Quantum Optical Master Equation (QOME) \cite{Breuer2006}. We conclude that the solutions to this equation are valid up to the same order of approximation as the solutions of the Redfield equation. The presented derivation shall be understood as the formalization of the secular approximation. We systematically remove terms of higher order in the coupling strength $g$ between system and bath from the formal solutions of the Redfield equation. As a result, we obtain the Quantum Optical Master Equation naturally, as we would by applying the secular approximation in the traditional, heuristic way.

It needs to be stated here that the validity of a Lindblad type equation in the Markov approximation has been shown already by Davies \cite{Davies1974}. The important difference here is that this work specifically shows the validity of the secular approximation generally applied to the Redfield Equation. It can therefore be seen as a specification of Davies work, where we prove explicitly that it is valid specifically for the secular approximation used to obtain the Quantum Optical Master Equation.

Also, it is important to note that this is only a statement about reduced dynamics at finite times. Especially for asymptotic states, the secular approximation can miss $\mathcal{O}(g^2)$-corrections \cite{Crowder_2024}.

\subsection{Comparison to Universal Lindblad} \label{sec:comp}
Eq.~\eqref{eq: QOME} is in Lindblad form, but its solutions are valid up to the same order of approximation in the bath coupling strength $g$ as the Redfield equation. The fact that there is an equation in Lindblad form, which is in the same equivalence class of approximations as the Redfield equation, is nothing new though. In fact, Rudner and Nathan have shown that there is another such equation, the ULE \cite{RudnerNathan}. The core difference to the result obtained here is that the ULE contains only one Lindblad operator $L_{\text{UL}}$ for each $q$.

It relates to Eq.~\eqref{eq: QOME} in that the sum over $\omega$ of the Lindblad operators in that equation equate the Universal Lindblad operator, so
\begin{equation}\label{eq: UL op}
    L_q^{(\text{UL})} = \sum_\omega \sqrt{\gamma_q(\omega)} \hat{S}_q(\omega).
\end{equation}

We give an analytical intuition for the better accuracy of the Quantum Optical 
Master Equation compared to the Universal Lindblad Equation. Following Ref.~\cite{RudnerNathan}, the Redfield equation in the interaction picture takes the form
\begin{align}
	\dot{\tilde{\rho}}(t) = \int_{-\infty}^\infty \dd t' \int_{-\infty}^\infty \dd s
	\mathcal{F}(t,s,t')[\tilde{\rho}]
\end{align}
with 
\begin{align}
	\mathcal{F}(t,s,t')[A] &= \gamma \theta(t - t') g(t-s) g(s-t')
	[\tilde{S}(t),A(\tilde{S}(t')] + \text{H.c.}
\end{align}
and
\begin{align}
	g(s) = \frac{1}{\sqrt{2\pi}} \int_{-\infty}^\infty \dd \nu \sqrt{J(\nu)}
	e^{-i\nu s}
\end{align}
with the bath spectral function $J$. We transform back to the Schr\"odinger picture. We get
\begin{align}
	\dot{\tilde{\rho}}(t) = i[H_S,\tilde{\rho}(t)] + R\tilde{\rho}(t)
\end{align}
with 
\begin{align}
	R\tilde{\rho}(t) &= 
	\int_{-\infty}^\infty \dd t' \int_{-\infty}^\infty \dd s
	U^\dagger(t) \mathcal{F}(t,s,t') U(t) \notag \\ &=
	\int_{-\infty}^\infty \dd t' \int_{-\infty}^\infty \dd s
	\left(S\tilde{\rho}(t) e^{iH_S(t-t')} S e^{-iH_S(t-t')} \right.
    - \left.
	\tilde{\rho}(t) e^{iH_S(t-t')} S e^{-iH_S(t-t')} S\right) + \text{H.c.}
\end{align}
We consider a two-level system with system
Hamiltonian $H_S = \epsilon \sigma_z + \Delta \sigma_x$ in the following. In its eigenbasis, the previous equation reads
\begin{align}
	R\tilde{\rho}(t) &= 
	\int_{-\infty}^\infty \dd t' \int_{-\infty}^\infty \dd s
	\Bigg[ \gamma \theta(t-t') g(t-s) g(s-t') \notag \\
	&\phantom{+} \sum_{a,b,c,d} e^{i\omega_{db}(t-t')}
	\bra{a}S\ket{c} \bra{d}S\ket{b} \ket{a}\bra{c}\rho(t)\ket{d}\bra{b} 
	\notag \\ &-
	\sum_{d,k,b} e^{i\omega_{dk}(t-t')}
	\bra{d}S\ket{k} \bra{k}S\ket{b} \rho(t) \ket{d}\bra{b}
	\Bigg] \notag \\ &+ \text{H.c.},
\end{align}
where $\ket{a}$ are the eigenstates of $H_S$. The choice of indices is for readability with respect to the next equation. We calculate the terms for the Redfield tensor by
\begin{align}
	R_{abcd} = \bra{a}(R\ket{c}\bra{d})\ket{b}.
\end{align}

We now consider for simplicity a system with $S = \sigma_z$ and $H_S = \sigma_x$
and $E_0 = 1$ and $E_1 = -1$ being the eigenenergies of $H_S$. Defining $\omega
= E_0 - E_1$ enables us to write down the Redfield superoperator with
$\mathcal{R}_{an + b,cn + d} = R_{abcd}$ explicitly as
\begin{align}
	\mathcal{R} = 
	\begin{pmatrix}
	-G(\omega) - G(-\omega) & 0 & 0 & G(\omega) + G(-\omega) \\
	0 & G(\omega) - G(-\omega) & G(\omega) - G(-\omega) & 0 \\
	0 & G(-\omega) - G(\omega) & G(-\omega) - G(\omega)  & 0 \\
	G(\omega) + G(-\omega) & 0 & 0 & -G(\omega) - G(-\omega) 
	\end{pmatrix}
\end{align}
with
\begin{align}
	G(\omega) &= 
	\int_{-\infty}^\infty \dd t' \int_{-\infty}^\infty \dd s
	\gamma \theta(t-t') g(t-s) g(s-t') e^{i\omega(t-t')}.
\end{align}

We describe the two approaches to bring the Redfield equation into Lindblad
form:
\begin{itemize}
	\item For the Quantum Optical Master Equation, we remove all the terms
	except
	$\mathcal{R}_{00},\mathcal{R}_{03},\mathcal{R}_{30},\mathcal{R}_{33}$ (from the incoherent poles) and $\mathcal{R}_{11},\mathcal{R}_{22}$ (from the coherent poles).
	\item For the Universal Lindblad equation, we replace the function $G$ by
	the function $F$ defined by \cite{RudnerNathan}
		\begin{align}
			F(\omega) &= 
			\int_{-\infty}^\infty \dd s' \int_{-\infty}^\infty \dd s 
			\gamma \theta(s-s') g(s-t) g(t-s) e^{i\omega(t-s')}.
		\end{align}
\end{itemize}

\begin{figure}
    \centering
    \small
	\begin{tikzpicture}[scale=2, align=center]
    	\node[] (1) {
            \footnotesize$ 
			\frac{\dd}{\dd t} \underline{\rho}(t) = 
			\begin{pmatrix}
			-\tilde{G}_+(\omega) & 0 & 0 & \tilde{G}_+(\omega) \\
			0 & i\omega + \tilde{G}_-(\omega) & \tilde{G}_-(\omega) & 0 \\
			0 & \tilde{G}_-(-\omega) & -i\omega + \tilde{G}_-(-\omega) & 0 \\
			\tilde{G}_+(\omega) & 0 & 0 & -\tilde{G}_+(\omega) 
			\end{pmatrix}
			\underline{\rho}(t) \quad \quad
		$};
    	\node[below right of=1, node distance=30ex, xshift = 7ex] (2)   {
            \footnotesize$
			\begin{pmatrix}
			-\tilde{G}_+(\omega) & 0 & 0 & \tilde{G}_+(\omega) \\
			0 & i\omega + \tilde{G}_-(\omega) & 0 & \\
			0 & 0 & -i\omega + \tilde{G}_-(-\omega)& 0 \\
			\tilde{G}_+(\omega) & 0 & 0 & -\tilde{G}_+(\omega)
			\end{pmatrix}
		$};
    	\node[below left of=1, node distance=30ex, xshift = -7ex] (3) {
            \footnotesize$
			\begin{pmatrix}
			-\tilde{F}_+(\omega) & 0 & 0 & \tilde{F}_+(\omega) \\
			0 & i\omega + \tilde{F}_-(\omega) & \tilde{F}_-(\omega) & 0 \\
			0 & \tilde{F}_-(-\omega) & -i\omega + \tilde{F}_-(-\omega) & 0 \\
			\tilde{F}_+(\omega) & 0 & 0 & -\tilde{F}_+(\omega) 
			\end{pmatrix}
		$};
		\draw[-stealth] (1) -- (2) node[right=4ex, midway]{QOME};
		\draw[-stealth] (1) -- (3) node[left=4ex, midway]{ULE};
	\end{tikzpicture}
	\caption{Comparison of the resulting superoperators of the Universal Lindblad equation (ULE) and the Quantum Optical master equation (QOME) with respect to the Redfield equation. We see that in the case of the Universal Lindblad equation, in addition to the matrix elements that can be compared to $\omega$, the other matrix elements are also changed. For simplicity, we defined $\tilde{G}_\pm(\omega) = G(\omega) \pm G(-\omega)$ and $\tilde{F}_\pm(\omega) = F(\omega) \pm F(-\omega)$.}
    \label{fig:compULQO}
\end{figure}

The changes made by the Quantum Optical Master Equation and the
Universal Lindblad Equation are visualized in Fig.~\ref{fig:compULQO}. We
see that for this simple case, the terms removed by application of the secular
aproximation are only the ones which are dominated by $\omega$ anyways. In the
Universal Lindblad case, the other matrix elements are changed as well.
Since in the weak coupling limit, $\omega$ is large compared to $G(\omega)$
or $F(\omega)$, changing $F$ to $G$ in terms containing $\omega$ has little
effect on the solution. However, in the case of the Universal Lindblad Equation,
even terms that are not dominated by $\omega$ are changed. This indicates that especially in cases where $G(\omega)$ becomes very small compared to $\omega$, the QOME should have higher accuracy than the ULE. 

We can furthermore note that the approximations for the QOME and the ULE differ in the approximation parameter. In the case of the QOME, we approximate $g^2$ with respect to $\omega$, hence we consider ${g^2}/{\omega_{\min}}$ as expansion parameter. In contrast, to obtain the ULE, an expansion in the parameter $g^2 \tau$ is used, where $\tau$ denotes the bath correlation time \cite{RudnerNathan}. Hence, as the bath correlation time decreases with increasing temperature, the ULE becomes more accurate for systems with higher bath temperatures. Consequently, the QOME shall outperform the ULE if the bath temperature is close to 0. In the following, we will demonstrate this fact numerically by presenting a simulation of a system which is more accurately described by QOME at lower temperatures, while the ordering reverses at higher temperatures.

\section{Numerical Examples} \label{sec:numerics}

In this section, we compare the performance of several well-known quantum master equations applied to a simple two-level system. Our primary focus is on contrasting the Quantum Optical Master Equation (QOME) with the Universal Lindblad Equation (ULE). We find that, for lower temperatures, QOME performs better than ULE in terms of accuracy, but the ordering reverses for higher temperatures. However, it is important to note that this result is limited to the setup considered here and should not be generalized as a broader proof for all systems. The goal of this comparison is to provide insights based on a controlled, simple example, rather than to offer a universal statement.

A similar comparison has been presented in Ref.~\cite{Davidovic2020completelypositive}, where the same master equations were applied, but to different test cases. In that work, the completely-positive Lindblad completion of Redfield (there called geometric-arithmetic master equation, GAME), which is similar to ULE-type dynamics, was compared to the exact solution for both a solvable V-type three-level system and a 25-spin-1/2 ferromagnetic chain with dipole-dipole coupling. In those examples, the QOME-like evolution (referred to as RWA in Ref.~\cite{Davidovic2020completelypositive}) was found to be less accurate, while ULE-like constructions showed improved agreement with Redfield dynamics.

In contrast, our simulations focus primarily on exploring how the various equations perform when different parameters and initial states are varied for the two-level system. As we will show, for the test case considered here, the order of accuracy of QOME and ULE depend on the system parameters, specifically the temperature.

The single-particle Hamiltonian describing the TLS is given by
\begin{equation}
\label{eq: singleQubitHam}
    \hat{H}_S = E \Bigl (\cos(\phi)\hat{\sigma}_z + \sin(\phi)\hat{\sigma}_x \Bigr ), 
\end{equation}
where $E$, $\phi$ are some constants describing the qubit energy and orientation in the Bloch sphere, respectively; also we have to consider the interaction with the environment, that we are going to model as an Ohmic or Jaynes-Cummings-like bath. Note that this is only valid in the Purcell regime, since the approach relies on perturbation theory.

Ohmic baths are ubiquitous in high frequency ($\gtrsim \SI{1}{\giga\hertz}$) spectroscopy accounting for Nyquist noise \cite{Krantz2019AQE},  the Ohmic spectral density function bath reads
\begin{equation}
    J(\omega)= \alpha \omega \frac{{\Omega}^2}{ \Omega^2 + \omega^2 },
\end{equation}
where $\alpha$ is some bath-amplitude parameter and the high frequency pseudo-Lorenz cutoff ${{\Omega}^2}/{( \Omega^2 + \omega^2) }$ prevents ultraviolet divergences for $\omega \gg \Omega$.

The Jaynes-Cummings-like (JC from now on) spectral density function
\begin{equation}
    J(\omega) = \frac{\alpha \omega \Omega ^4}{(\omega^2 - \Omega^2)^2+4\pi^2 \Gamma^2 \omega^2\Omega^2}
\end{equation}
as discussed in~\cite{Garg83, wilhelm2007superconducting}, can be derived by assuming that an oscillator is coupled via another oscillator of frequency $\Omega$ to infinite many oscillators having an Ohmic spectrum $J(\omega)=\Gamma \omega \Theta(\omega-\omega_c)$ up to some cutoff frequency $\omega_c \gg \Omega$. 
This spectrum is known to mimic the “damped” behavior of an open Jaynes-Cummings model or a spin-boson model.

As said at the beginning of this Section, we now want to compare the QOME with the Universal Lindblad equation \cite{RudnerNathan}. The first thing to notice is that the ULE has only one Lindblad operator, while the QOME has multiple Lindblad operators, their number depending on the degeneracy of the system Hamiltonian. We are going to show that for the chosen parameters here, the numerical results for QOME are much closer to the Redfield-induced dynamics than the ULE.

\begin{figure}[ht!]
\centering
    \includegraphics[width=\textwidth]{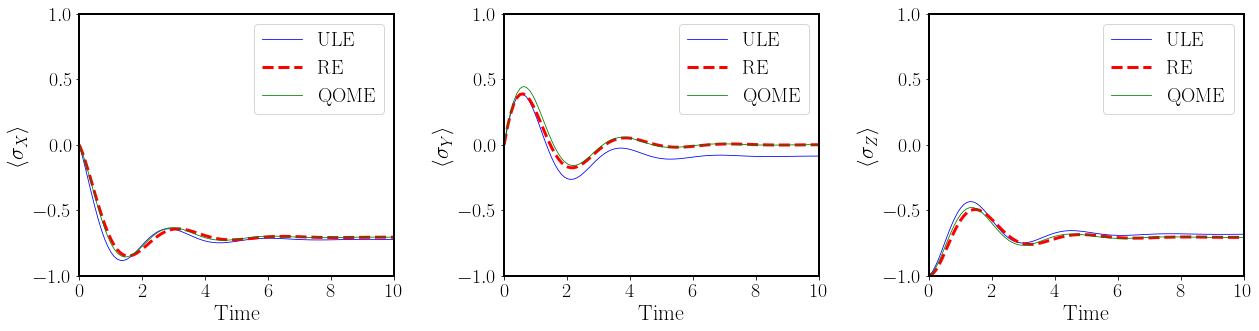}
\caption{Comparison of the time dependent expectation values of the Pauli operators $\langle \hat{\sigma}_i\rangle$ after evolving a TLS coupled to a JC bath with the ULE, Redfield Equation (RE) and QOME respectively indicated in blue, red and green. Setting $E = 1$, so that  $t_{\text{final}} = 10/E$, the parameters used to simulate the plotted evolution are as follows: $\alpha = 0.2$, $T = 0.05$, $\theta = {\pi}/{2}$, $\phi = {\pi}/{4}$, $\varphi = {\pi}/{2}$, $\Omega = 50$.}
\label{Fig: UL-QO dynamics}
\end{figure}

As an example, Fig. \ref{Fig: UL-QO dynamics} showcases the different time dependent expectation values of the Pauli operators after evolving with the ULE (blue line), the QOME (green line) and the Redfield equation (red line). QOME and RE are completely overlapping and therefore undistinguishable, highlithing the difference with the ULE. However, the comparison here is only for one set of parameters and can therefore not serve as a general indicator.

We now want to setup a more systematic comparison of the ULE and the QOME. In order to do that we keep the Redfield Equation as a benchmark, fix a time at which we measure the distance to the benchmark, and vary the different parameters related to the TLS and the bath. 
We start by defining the trace distance at a given time $t_f$ between two different evolutions as the trace distance between the two evolved density matrices at that given time. For example, we want to compare the propagation induced by the ULE with respect to the Redfield equation, we thus compute

\begin{equation}
     \mathcal{D}_\text{UL-RE}(t_f)=T (\hat{\rho}_{\text{UL}}(t_f), \hat{\rho}_{\text{RE}}(t_f))
\end{equation}
where the trace distance $T(\cdot)$ is defined as 
\begin{equation}
    T(\hat{\rho}, \hat{\sigma})=\dfrac{1}{2}\Tr{\sqrt{(\hat{\rho}-\hat{\sigma})^{\dagger}(\hat{\rho}-\hat{\sigma})}}
\end{equation}
see \textit{e.g.}~\cite{Breuer2006}, and analogously we define
   \begin{equation}
     \mathcal{D}_\text{QO-RE}(t_f)=T (\hat{\rho}_{\text{QO}}(t_f), \hat{\rho}_{\text{RE}}(t_f)).
\end{equation}

\begin{figure*}[ht!]  
    \centering
    \begin{minipage}{0.49\textwidth}
        \centering
        \includegraphics[width=\textwidth]{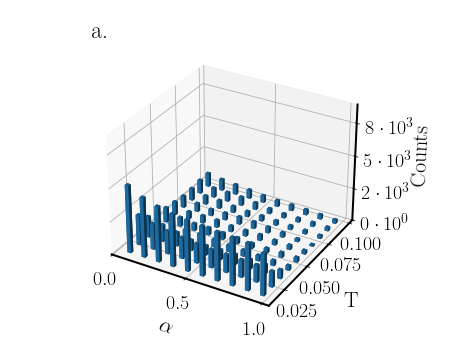}  
    \end{minipage}\hfill
    \begin{minipage}{0.49\textwidth}
        \centering
        \includegraphics[width=\linewidth]{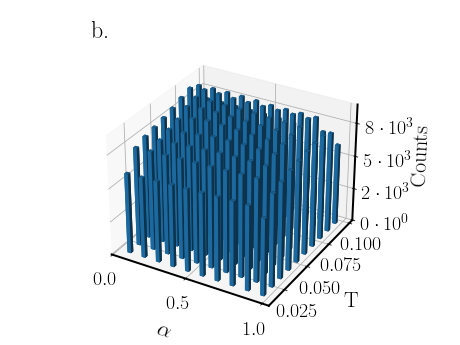}  
    \end{minipage}

    \vspace{0.1cm}  

    \begin{minipage}{0.49\textwidth}
        \centering
        \includegraphics[width=\linewidth]{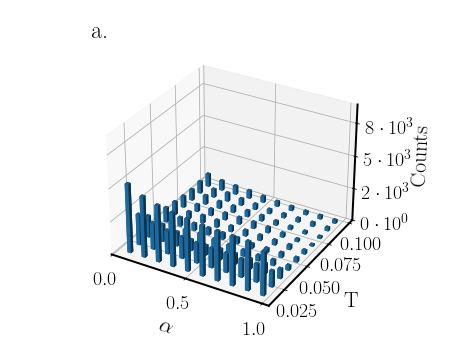}  
    \end{minipage}\hfill
    \begin{minipage}{0.49\textwidth}
        \centering
        \includegraphics[width=\linewidth]{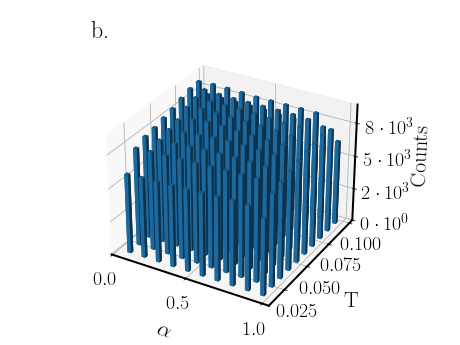}  
    \end{minipage}
    \caption{Histograms showing the number of instances whose final states (at a given time $t_f=10/E_1$) have a trace distance smaller than a certain value $\mathcal{D}_{\text{thres}}=0.01$ as a function of the coupling to the bath $\alpha$ and the temperature $T$. The parameters $\alpha$ and $T$ are linearly sampled within the intervals in Table~\ref{Tab: ParamsLin}, while the mixing angle and the Bloch sphere angles are randomly chosen within the ranges in Table~\ref{Tab: ParamsInit}. Plots a. and c.: $\mathcal{D}_{\text{UL-RE}}(t_f)< 0.01$ for an Ohmic and a JC bath respectively. Plots b. and d.: $\mathcal{D}_{\text{QO-RE}}(t_f)< 0.01$ for an Ohmic and a JC bath respectively.}
    \label{Fig: 3d histograms}
\end{figure*}

\begin{figure*}[ht!]  
    \centering
    \begin{minipage}{0.49\textwidth}
        \centering
        \includegraphics[width=\textwidth]{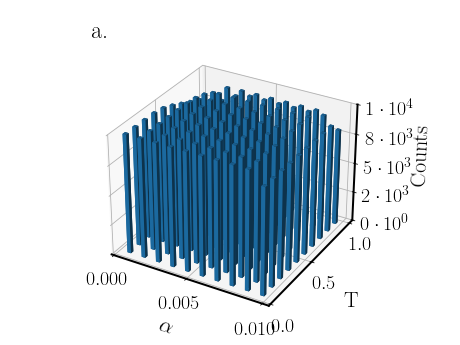}  
    \end{minipage}\hfill
    \begin{minipage}{0.49\textwidth}
        \centering
        \includegraphics[width=\linewidth]{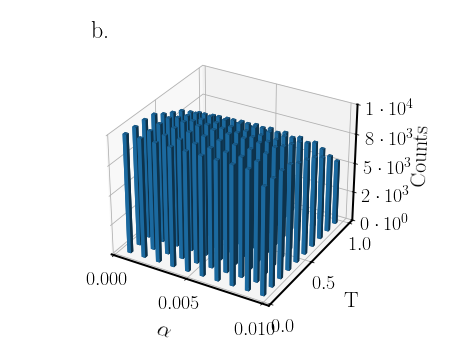}  
    \end{minipage}

    \vspace{0.1cm}  

    \caption{Histograms showing the number of instances whose final states (at a given time $t_f=10/E_1$) have a trace distance smaller than a certain value $\mathcal{D}_{\text{thres}}=0.01$ as a function of the coupling to the bath $\alpha$ and the temperature $T$. The parameters $\alpha$ and $T$ are linearly sampled within the intervals $\alpha \in [0.0, 0.01]$ and $T \in [0., 1.]$, while the mixing angle and the Bloch sphere angles are randomly chosen within the ranges in Table~\ref{Tab: ParamsInit}. Plot a. shows again the count of instances for which $\mathcal{D}_{\text{UL-RE}}(t_f)< 0.01$, while plot b. shows the amount of instances for which $\mathcal{D}_{\text{QO-RE}}(t_f)< 0.01$. Both plots are the results of coupling to a JC bath.}
    \label{Fig: 3d histograms high T}
\end{figure*}

To compare the accuracy of the QOME and the UL with respect to the Redfield Equation, we pick 100 instances of the parameters $\phi$, $\theta$, $\varphi$, randomly defining the initial state and the mixing angle of the Hamiltonian, whitin the ranges in Table~\ref{Tab: ParamsInit}; and we sample 100 linearly spaced values for the coupling to the bath $\alpha$ and the temperature $T$ whitin the ranges in Table~\ref{Tab: ParamsLin}. The initial state $\rho(t=0)=\ketbra{\psi_0}{\psi_0}$ is given by 
\begin{equation}
    \ket{\psi_0} = \cos{(\theta/2)}\ket{0}+e^{i\varphi}\sin{(\theta/2)\ket{1}}
\end{equation}
and contains the parameters $\theta$ and $ \varphi$. 
The histograms in Fig. \ref{Fig: 3d histograms} show how many instances evolved with the ULE and the QOME end up having  trace distance smaller than a certain threshold $\mathcal{D}_{\text{thres}}=0.01$. The height of the bars in the pair of plots on the left indicates the number of simulations for which $\mathcal{D}_{\text{UL-RE}}(t_f)<\mathcal{D}_{\text{thres}}$; analogously, on the right are plotted the counts of density matrices evolved with the QOME satysfying $\mathcal{D}_{\text{UL-RE}}(t_f)<\mathcal{D}_{\text{thres}}$. The type of bath does not seem to play a role since the upper pair, featuring an Ohmic spectral density function, looks pretty much the same as the bottom pair, where the spectral density function is JC. We notice that the accuracy of the QOME with respect to the Redfield Equation is on average higher than that of the ULE. However, this conclusion is strongly dependent on the temperature regime. In particular, at lower temperatures, the QOME clearly outperforms the ULE across a broad range of coupling strengths. In contrast, as the temperature increases, this trend reverses and the ULE becomes the more accurate description, see Fig.~\ref{Fig: 3d histograms high T}. Only when the bath coupling $\alpha$ and temperature $T$ are much smaller than the qubit energy, so $E_1\gg \alpha,\, T$, the solutions become close in accuracy. In these cases, we expect coupling to the environment to be small enough that the effect of the Lindblad operators becomes negligible. At present, the underlying reason why one approximation outperforms the other in different temperature regimes is not fully understood. In particular, the mechanisms responsible for the improved performance of the QOME at low temperatures, and the crossover to ULE dominance at higher temperatures, remain unclear and warrant further investigation.

Importantly, these conclusions should not be regarded as general statements about the relative performance of the two master equations. Rather, they are restricted to the specific and highly simplified scenario considered here. Different system architectures, coupling structures, or bath models may lead to qualitatively different behavior.

As expected from the theoretical considerations in section \ref{sec:comp}, this numerical analysis indicates that while the QOME is preferable in low-temperature regimes within this model, the ULE can provide a better approximation at higher temperatures, underscoring both the parameter dependence of the results, their limited scope of applicability, and the need for a deeper theoretical understanding.

\begin{table}
\centering
\caption{Random parameters of the TLS Hamiltonian and initial state}
\label{Tab: ParamsInit}
\begin{tabular}{||c | c | c||} 
 \hline
 Parameters & Min. Value & Max. Value \\ [0.5ex] 
 \hline\hline
 $\phi$ & 0 & $\pi/4$ \\ 
 \hline
 $\theta$ & 0 & $\pi$ \\
 \hline
 $\varphi$ & 0 & $2\pi$  \\ 
 \hline
\end{tabular}
\end{table}

\begin{table}[htbp]
\centering
\caption{Linearly sampled coupling to the bath and temperature}
\label{Tab: ParamsLin}
\begin{tabular}{||c | c | c||} 
 \hline
 Parameters & Min. Value & Max. Value \\ [0.5ex] 
 \hline\hline
 $\alpha$ & 0.1 & $1$ \\ 
 \hline
 $T$ & 0.01 & 0.1 \\

 \hline
\end{tabular}
\end{table}

\section{Conclusions}
In this work, we have shown a way to formally derive the Quantum Optical Master Equation from the Redfield Equation. This systematic way of deriving the QOME only discards terms in the solution of the master equation that are of the same or higher order than those already neglected to derive the Redfield equation. The QOME therefore belongs to the same equivalence class of Markov approximations as the Redfield Equation. A similar procedure has already been done by Rudner and Nathan, who showed that the Universal Lindblad Equation also belongs to this equivalence class \cite{RudnerNathan}. 

This result not only formally justifies the secular approximation that is usually done in a very informal way by just neglecting "fast oscillating terms", which in itself is of
high analytical interest. Additionally, the formalism presented here could serve as a starting point for treatment of higher orders of the Nakaajima-Zwanzig equation, providing a way to generalize the secular approximation beyond the Redfield equation. 

We then provided a theoretical comparison of the QOME and the ULE, indicating that the accuracy of the ULE compared to the QOME depends on the bath temperature, where higher temperatures result in higher accuracy of the ULE compared to the QOME.
    
To test this hypothesis, we applied both the QOME
and the ULE to solve single-qubit systems with varied parameters,
and we compared the solutions to the solution of the Redfield Equation in certain cases where the latter retains valid results. For the system we considered, we found that the accuracy of the QOME compared to the ULE depends on the parameters of the system. Specifically, we found that for higher temperatures, ULE yielded more accurate results than QOME, while for lower temperatures, the reverse was the case. It remains an open question how this observation carries over to more complex systems. A natural continuation of this work would be to do numerical testing on more complex models to find more specific indicators for when the ordering reverses. 

Furthermore, one caveat of the QOME is that it
requires knowledge of the system
eigenstates, while the ULE does not. A useful prospect
for further work would be to see if the Lindblad operators of the Quantum
Optical Master Equation can be derived without knowledge of the system
eigenstates in a similar way as the Lindblad operator of the ULE can.

In conclusion, we have shown a formal way to derive the QOME from the Redfield Equation for any open quantum system. With that, we have shown that the QOME yields solutions that are valid up to the same order of approximation than those of the Redfield equation, a result which is of great analytical and practical significance.

\newpage
\subsection*{\textbf{Acknowledgements}}

We wish to acknowledge useful discussions with J\"urgen Stockburger and Frederik Nathan. The simulations have been performed using the Python package QuTiP~\cite{johansson2012qutip}.

Verbundprojekt: Digital-Analoge Quantencomputer (DAQC) – Teilvorhaben: DAQC Kontrolle, Kalibrierung und Charakterisierung, BMFTR, FKZ: 13N1568.

Impuls- und Vernetzungsfonds, Vorhaben QRUISE, Helmholtz-Gemeinschaft Deutscher Forschungszentren (HGF), FKZ: HVF-0096.

Diamond spin-photon-based quantum computer (Spinning), BMFTR, FKZ: 13N16210.


\bibliography{QOME}

\end{document}